\documentclass[]{pasj02} 

\jyear{2024}
\Received{}
\Accepted{}

\begin{document}

\title{Decadal brightening in the northeastern non-thermal filament of Cassiopeia A}

\author{
Nao \textsc{Kominato}\altaffilmark{1}\altemailmark\orcid{0000-0001-8335-1057} \email{kominato.nao@rikkyo.ac.jp}, 
Toshiki \textsc{Sato}\altaffilmark{2}\orcid{0000-0001-9267-1693},  
Ryota \textsc{Hayakawa}\altaffilmark{1,3}\orcid{0000-0002-3752-0048}, 
Yusuke \textsc{Sakai}\altaffilmark{1} \orcid{0000-0002-5809-3516}, 
and
Shinya \textsc{Yamada}\altaffilmark{1}\orcid{0000-0003-4808-893X}
}
\altaffiltext{1}{Department of Physics, Rikkyo University, 3-34-1 Nishi Ikebukuro, Toshima-ku, Tokyo 171-8501, Japan}
\altaffiltext{2}{Department of Physics, School of Science and Technology, Meiji University, 1-1-1 Higashi Mita, Tama-ku, Kawasaki, Kanagawa 214-8571, Japan}
\altaffiltext{3}{International Center for Quantum-field Measurement Systems for Studies of the Universe and Particles (QUP), KEK, 1-1 Oho, Tsukuba, Ibaraki 305-0801, Japan}

\footnotetext[$\dag$]{Present address: }



\maketitle

\begin{abstract}
We present the decadal brightening of non-thermal emission flux in the northeastern filament of the young supernova remnant Cassiopeia A (Cas A), which highlights dynamic processes in the forward shock. This filament, characterized by the highest particle acceleration rate among Cas A's outer shells, offers an exceptional opportunity to investigate underlying astrophysical mechanisms. Since 2000, the non-thermal flux has increased by several tens of percent before plateauing, while the spectral shape has remained largely unchanged. Over the past two decades, the filament's morphology has evolved significantly, splitting into two distinct sections. Detailed analysis reveals contrasting behaviors: one section shows a flux increase followed by saturation, while the other maintains a steady flux. These differences likely arise from nonlinear effects, including magnetohydrodynamic interactions influenced by magnetic field orientation, interactions with surrounding material, and complex fluid dynamics associated with young supernova remnants. The localized evolution of this filament, captured with high spatial resolution, provides critical insights into the temporal dynamics of non-thermal particles and the generation of cosmic rays from asymmetric supernova explosions.
\end{abstract}


\section{Introduction}\label{sec:intro}

Diffusive shock acceleration in supernova remnants (SNRs) is a leading theory for explaining the origins of galactic cosmic rays.
Observational tests of this theory have been conducted using numerous young shell-type SNRs, which are known to emit non-thermal X-ray radiation. Decades of X-ray imaging and spectroscopic studies by missions such as \textit{ASCA}(\cite{Koyama:1995aa}), \textit{XMM-Newton}, and \textit{Chandra} have provided substantial evidence supporting this mechanism. It is now widely accepted that non-thermal X-ray emissions from these objects are primarily due to synchrotron radiation produced by relativistic electrons (\cite{2000ApJ...528L.109H}).

Despite significant evidence for particle acceleration in SNRs, several key questions remain unresolved. These include the mechanisms initiating seed particle injection, the factors sustaining their acceleration, and the dynamics governing their escape or further acceleration. The complexity of these processes is influenced by various factors, including magnetic field strength and orientation, interactions with the surrounding medium, the maximum energy achievable by accelerated particles, the remnant's evolutionary speed, and anisotropies in the ejected material. High spatial resolution studies are essential for advancing our understanding of these processes and addressing these longstanding challenges.

\begin{longtable}{ccccccc}
  \caption{Comprehensive log of Chandra observations for Cas~A. }\label{obstable}  
\hline\noalign{\vskip3pt} 
  Obs. ID &Obs. Start & Exposure &Detector& RA&Dec& Roll\\ 
{}  &(dd mm yyyy) &(ks) & & (deg) &(deg) &(deg)\\   [2pt] 
\hline
\endfirsthead      
\hline\noalign{\vskip3pt} 
  Name & Value1 & Value2 & Value3 & & & \\  [2pt] 
\hline\noalign{\vskip3pt} 
\endhead
\hline\noalign{\vskip3pt} 
\endfoot
\hline\noalign{\vskip3pt} 
\multicolumn{2}{@{}l@{}}{\hbox to0pt{\parbox{160mm}{\footnotesize
\hangindent6pt\noindent
}\hss}} 
\endlastfoot 
114 & 30 Jan 2000 & 49.93 & ACIS-S & 350.9159 & 58.7926 & 323.3801 \\ \hline
1952 & 06 Feb 2002 & 49.66 & ACIS-S & 350.9137 & 58.7923 & 323.3835 \\ \hline
5196 & 08 Feb 2004 & 49.53 & ACIS-S & 350.9129 & 58.7933 & 325.5035 \\
4638 & 14 Apr 2004 & 164.53 & ACIS-S & 350.9196 & 58.8365 & 40.3327 \\
5319 & 18 Apr 2004 & 42.26 & ACIS-S & 350.9127 & 58.8411 & 49.7698 \\
4636 & 20 Apr 2004 & 143.48 & ACIS-S & 350.9129 & 58.8412 & 49.7698 \\
4637 & 22 Apr 2004 & 163.49 & ACIS-S & 350.9131 & 58.8414 & 49.7665 \\
4639 & 25 Apr 2004 & 79.05 & ACIS-S & 350.9132 & 58.8415 & 49.7666 \\
4634 & 28 Apr 2004 & 148.63 & ACIS-S & 350.9047 & 58.8455 & 59.2239 \\
4635 & 01 May 2004 & 135.04 & ACIS-S & 350.9048 & 58.8455 & 59.2237 \\
5320 & 05 May 2004 & 54.37 & ACIS-S & 350.8988 & 58.8480 & 65.1350 \\ \hline
9117 & 05 Dec 2007 & 24.84 & ACIS-S & 350.8752 & 58.7846 & 278.1321 \\
9773 & 08 Dec 2007 & 24.84 & ACIS-S & 350.8753 & 58.7844 & 278.1318 \\ \hline
10935 & 02 Nov 2009 & 23.26 & ACIS-S & 350.8329 & 58.7868 & 239.6794 \\
12020 & 03 Nov 2009 & 22.38 & ACIS-S & 350.8330 & 58.7871 & 239.6796 \\ \hline
10936 & 31 Oct 2010 & 32.24 & ACIS-S & 350.8299 & 58.7888 & 236.4820 \\
13177 & 02 Nov 2010 & 17.24 & ACIS-S & 350.8298 & 58.7892 & 236.4818 \\ \hline
14229 & 15 May 2012 & 49.09 & ACIS-S & 350.8878 & 58.8478 & 75.4420 \\ \hline
14480 & 20 May 2013 & 48.77 & ACIS-S & 350.8895 & 58.8444 & 75.1402 \\ \hline
14481 & 12 May 2014 & 49.42 & ACIS-S & 350.8901 & 58.8423 & 75.1374 \\ \hline
14482 & 30 Apr 2015 & 49.42 & ACIS-S & 350.9080 & 58.8554 & 67.1266 \\ \hline
19903 & 20 Oct 2016 & 24.65 & ACIS-S & 350.8156 & 58.7909 & 214.1956 \\
18344 & 21 Oct 2016 & 25.75 & ACIS-S & 350.8149 & 58.7903 & 214.1979 \\ \hline
19604 & 16 May 2017 & 49.53 & ACIS-S & 350.8910 & 58.8560 & 76.5775 \\ \hline
19605 & 15 May 2018 & 49.42 & ACIS-S & 350.8927 & 58.8557 & 75.2332 \\ \hline
19606 & 13 May 2019 & 49.42 & ACIS-S & 350.8854 & 58.8559 & 75.1398 \\ \hline
\end{longtable}

 \begin{figure}[tbhp]
 \includegraphics[width=0.47\textwidth]{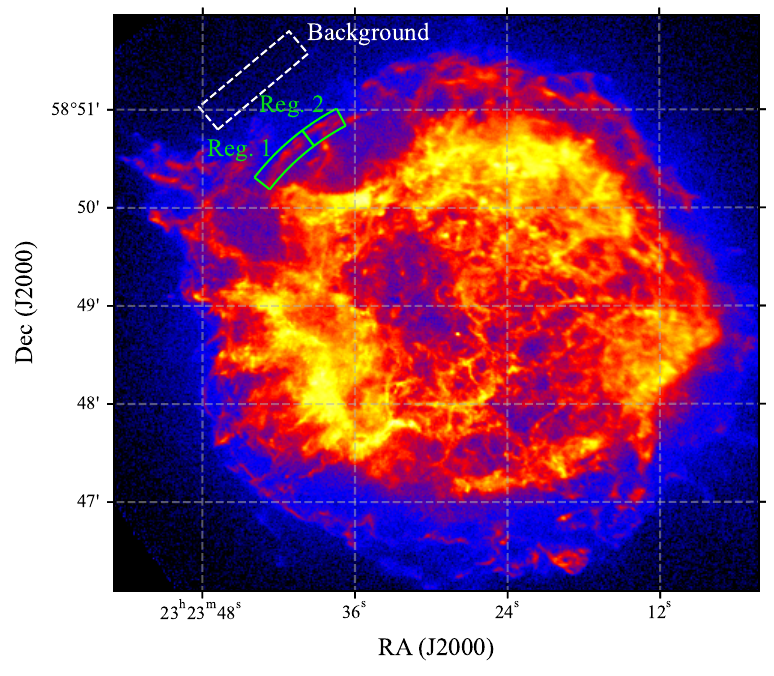}
 \caption{
High-resolution images of Cas~A in the 0.5--7.0 keV energy band captured by \textit{Chandra}, illustrating the intricate structures within the supernova remnant. 
The images highlight the two filamentary structures in the northeast region, outlined with solid lines (Region 1and Region2), 
with the background region indicated by dashed lines.
{Alt text: Overall view of CasA.} }
 \label{CasAfil}
 \end{figure}
 
Cassiopeia A (Cas~A) is one of the most luminous young SNRs in the X-ray domain. Observational studies over the past decade have revealed significant spectral variations, indicating the co-evolution of thermal and non-thermal particles within the remnant (\cite{Patnaude_2009, Uchiyama:2007aa}). Observations with \textit{NuSTAR} have been particularly revealing, demonstrating that a substantial fraction of the hard X-rays (greater than 10 keV) from Cas A originates from non-thermal processes. This finding is further supported by the detection of GeV emissions by Fermi and subsequent observations in the GeV-TeV spectrum by Fermi, MAGIC, and VERITAS (\cite{2009ApJ...693..713S}). Additionally, \textit{Chandra}'s high spatial resolution enables a detailed examination of the time evolution of non-thermal filaments. The highly asymmetric distribution of ejecta within Cas A has been extensively documented across multiple wavelengths (\cite{2001AJ....122.2644F, 2006ApJ...636..859F, 2010ApJ...725.2038D, 2014ApJ...785....7D, 2010ApJ...725.2059I, 2012ApJ...757..126I, 2014MNRAS.441.2996A, 2014ApJ...789..138P, 2015Sci...347..526M, 2016ApJ...818...17F}).

Recent X-ray studies using \textit{Chandra} have reported the detection of titanium (Ti) within Cas A (\cite{2014Natur.506..339G, 2017ApJ...834...19G, 2022arXiv220103753I}), along with evidence of inverted stratification that suggests an asymmetric supernova explosion (\cite{2021Natur.592..537S, 2021ApJ...912..131T}). As an alternative approach, deconvolution techniques have been explored to enhance morphological analysis, restoring fine angular resolution even in off-axis regions (\cite{Sakai_2024, RL_2023}). Localized analyses with high spatial resolution have enabled detailed investigations into the complex structures of Cas A. Among these, the northeastern filament stands out as the most intriguing region for particle acceleration, characterized by an estimated highly amplified magnetic field and efficient cosmic-ray diffusion nearing the Bohm limit (\cite{Stage:2006aa}). This paper focuses on the localized variations in non-thermal emission within the northeastern filament over the past decades. In areas of pronounced brightness variation, the flux has increased by approximately 80\% over 12 years before saturating. Details of the observations are presented in Section 2, while the analyses and discussions are covered in Sections 3 and 4, respectively.

\section{Observation} \label{sec:obs}

Cas~A has been observed with the Advanced CCD Imaging Spectrometer (ACIS) aboard the Chandra X-ray Observatory for detailed studies spanning over two decades, commencing in 1999. The observational data are summarized in Table \ref{obstable}. For our study, we compiled and analyzed available ACIS data until 2019, amounting to a total exposure time of approximately 1.5 Ms. Due to variability in the pointing axis across different observation sessions, the positional dependencies of the point spread function affect angular resolution. However, as this study focuses on spectral variations in two specific regions, the impact of these effects is minimal.

To enhance photon statistics, observations conducted within the same calendar year were combined into single datasets using the {\tt merge\_obs} command in the Chandra Interactive Analysis of Observations (CIAO) software suite. The data were then reprocessed using the {\tt chandra\_repro} command in CIAO 4.13, with calibration files updated to CALDB 4.9.5. For the highest accuracy in spatial analysis, the images were refined using the {\tt wcs\_update} command, utilizing the Central Compact Object within Cas~A as a reference point, since it is the only common point source across all observations. The resulting image, showing the 0.2-10.0 keV band, is presented in Figure \ref{CasAfil}.

 \begin{figure*}[tbhp]
 \begin{center}
 \includegraphics[width=0.98\textwidth]{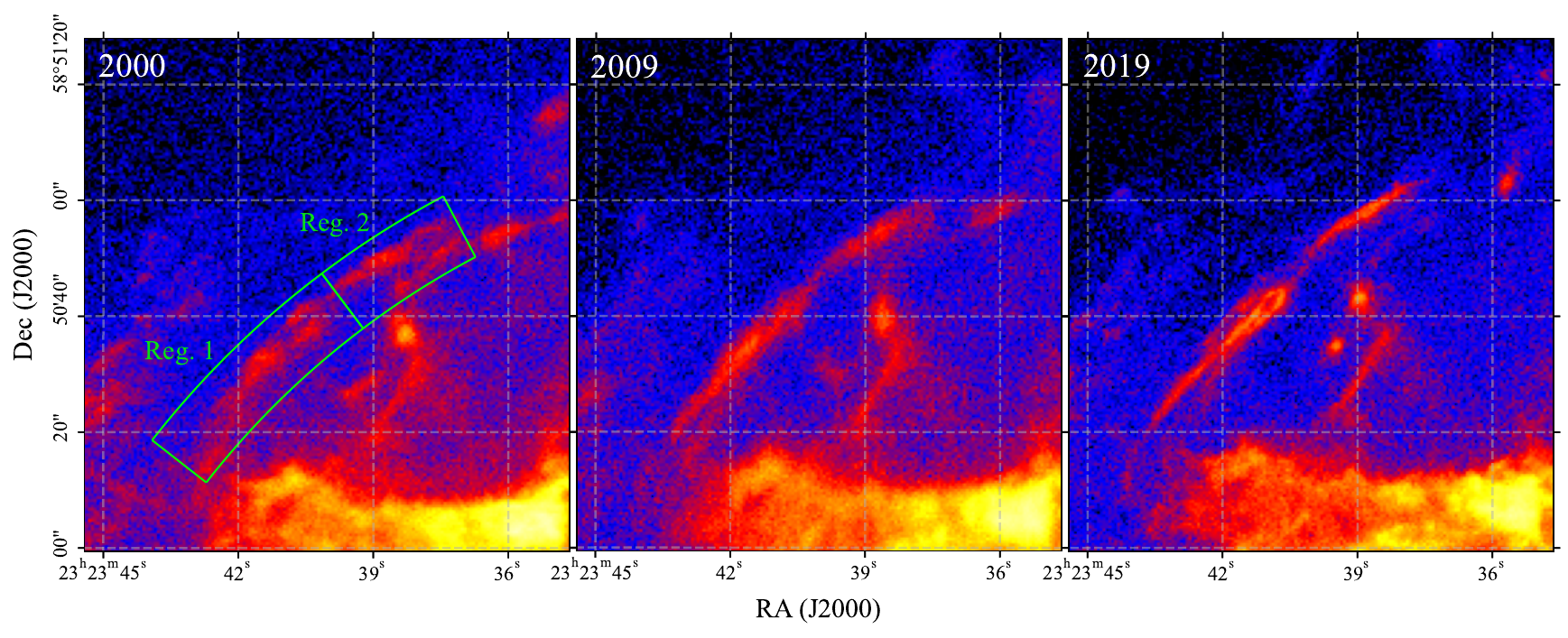}
 \end{center}
 \caption{
Sequential observations of the northeast filaments in Cas~A within the 0.5--7.0 keV energy band: this series of images displays the evolving structure of the northeast filaments over a span of fifteen years, captured in 2000, 2009, and 2019, respectively. Each image, progressing from left to right, highlights the dynamic changes occurring within this region of CasA, providing insights into the remnant's ongoing development and the physical processes driving these transformations.
{Alt text: Expanded view of the analysis area for the years 2000, 2009, and 2019.} } 
 \label{compare}
 \end{figure*}
 
Previous investigations into the shell of Cas~A have presented its properties, especially highlighting a notable trend in the fluxes of non-thermal emission. These variations in flux are hypothesized to follow a complex pattern rather than a simple monotonic progression. For instance, an increased cooling rate due to amplified magnetic fields or energized electrons  indicates a corresponding energy loss among the particles. This energy loss process is expected to reach a critical point where the decrease in energy becomes significant enough to stabilize the gas, resulting in a plateau. This naive scenario has driven our focused studies on the flux variations within the northeastern filament of Cas~A. This region is pivotal for understanding the detailed dynamical processes that govern particle acceleration.

\section{Data Analysis} \label{sec:ana}
\subsection{Spatial Temporal Variation}

The dynamic nature of Cas~A is underscored by the rapid expansion of its shock front, 
which propagates at an approximate velocity of $\sim$4500 km/s. This velocity translates to a motion of $\sim$0.28 arcsec/year for the northeastern filament, 
necessitating the adjustment of analysis regions to account for this movement (\cite{Patnaude_2009}). 
 Since the filament's displacement is measured to be roughly 0.6 pixels on the image per year, the coordinates of the analysis area are recalibrated for each observation date to maintain accuracy. 
This methodological adjustment ensures that the analysis remains consistent and reliable over time, capturing the evolving dynamics of Cas~A with a certain precision.

The morphological transformation of the northeastern filament, as captured in Figure \ref{compare}, presents the complex processes at play within CasA. 
In 2000, the filament exhibited an arc-like shape, typical of such remnants. 
However, a gradual transformation is observed through to 2019, with the filament not only straightening but also experiencing a notable thinning and brightening. 
This change results in the filament splitting into two distinct parts. 
A background region is selected at the detector's field of view edge and outside the SNR across all observed years, 
thereby without the need for adjustment like the main analysis region. 
The flux from this background region, contributing a few percent to the overall analysis,  
remains static, serving as a reference against the dynamic changes within the analysis regions. 

\subsection{Spectral Analysis}

The spectral analysis is conducted with the selection of specific regions within the north eastern filament to accurately estimate the fluxes of non-thermal emission. 
This selection was critical due to the filament's gradual morphological changes, specifically its splitting. 
To this end, three distinct regions were chosen for detailed study: the left side (Region 1), the right side (Region 2), and the entire filament (encompassing both Region 1 and Region 2). 
The delineation of these regions, as depicted in Figure~\ref{CasAfil}, was conducted, taking into account factors such as proper motion to ensure the reliability of our spectral fits.

We considered three different methods to quantify the data, ensuring a balance between robustness and conciseness. The first method estimated the background using two adjacent regions, approximating the thermal component from actual data and subtracting it from the total spectra. While this approach revealed consistent flux variations, it introduced uncertainties due to background selection. The second method utilized a background region outside the supernova remnant and applied a broad-band spectral fit, incorporating both thermal and non-thermal components. Although similar flux variations were observed, the non-thermal parameters were slightly affected by variations in the thermal plasma model. The third method focused on the 4.2--6.0 keV band, where the non-thermal component dominates, providing a straightforward means of quantifying its increase. Since the first and second methods indicated a non-thermal flux increase exceeding 10\%, surpassing systematic uncertainties, and the third method yielded consistent results, we adopted the third method in our analysis.
The first and second methods are described in Appendix 2.

By fitting the spectrum over a broad energy range using the thermal plasma model and the synchrotron radiation model, it was found that above 4 keV, the contribution from the thermal plasma component is sufficiently small compared to that of the synchrotron radiation component.
In terms of signal detection, the count rate across the entire region of interest, including the background, was approximately 1~c/s. This rate is sufficiently low to avoid issues related to pileup, thereby ensuring the integrity of the observed signal. The selection of the background region was equally rigorous, chosen from areas outside the remnant's shell to avoid any stray light contribution from the remnant. Notably, the background level within this energy range was approximately 5\% of the signal, allowing for a clear differentiation between the signal and background noise.

\begin{figure}[h]
 \includegraphics[width=0.45\textwidth]{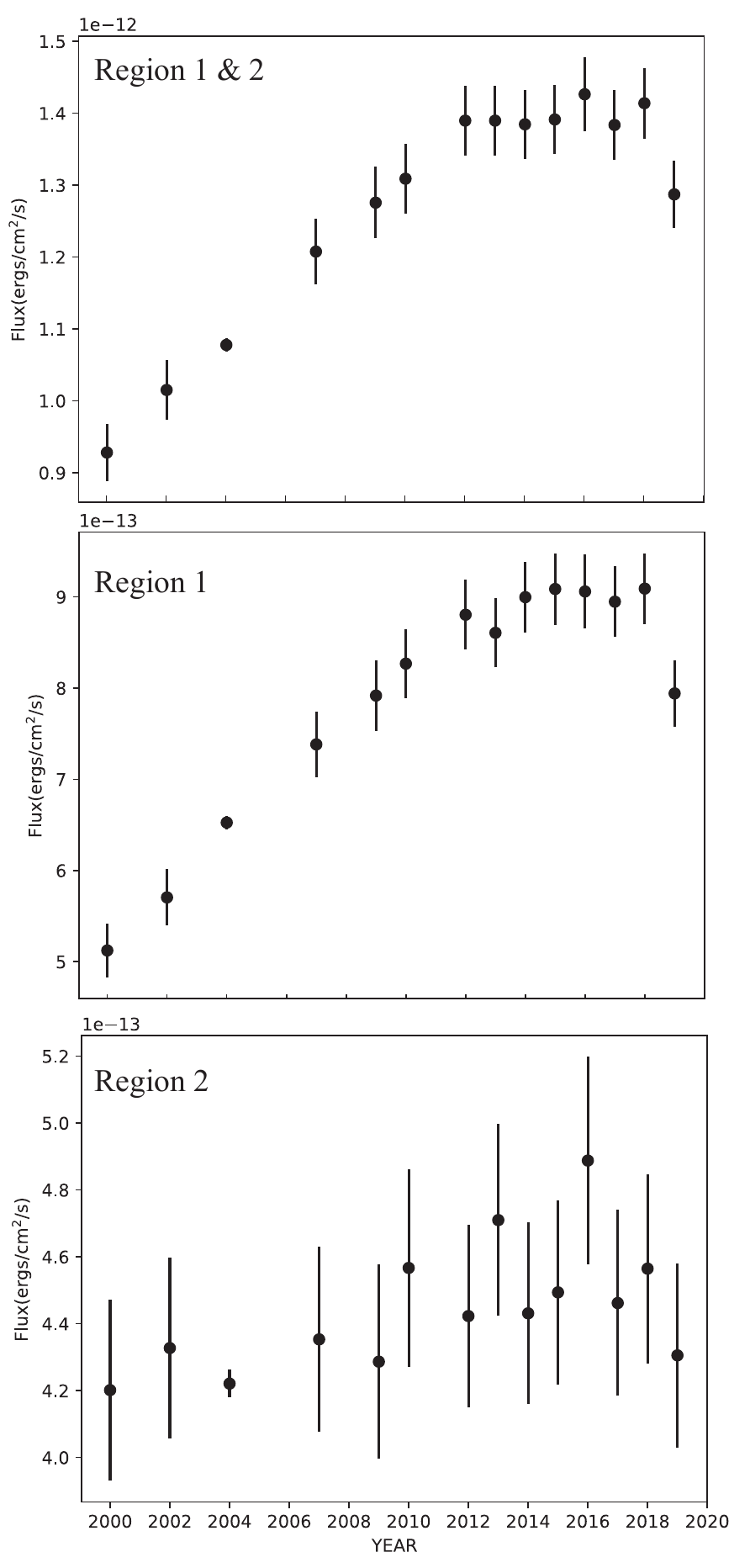}
 \caption{Flux variations obtained by fitting 4.2--6 keV range within the north eastern filament of Cas~ A over nearly two decades. The top, middle, and bottom panel displays the combined flux changes observed in Region 1 and 2, Region 1,  and Region 2, respectively.
 {Alt text: Three scatter plots graphs. In all graphs, the x axis  shows observation year from 2000 to 2020. In the top panel, the y axis shows flux in Region1 and 2 from 0.9e-12 to 1.5e-12 ergs per square centimeter per second. In the middle panel, the y axis shows flux in Region1 from 4.5e-13 to 10e-13 ergs per square centimeter per second.In the middle panel, the y axis shows flux in Region2 from 4.0e-13 to 5.2e-13 ergs per square centimeter per second.} } 
 \label{flux}
 \end{figure}

To quantify the changes in the spectral characteristics of the selected regions, we employed a simple model: a single power-law model with interstellar absorption. The known column density of the region approximately $1.0 \times 10^{22}$ cm$^{-2}$,so we set it at this value. The fitting results, illustrated in Figure~\ref{flux} and Table~\ref{results}, were statistically satisfactory across all regions, underscoring the model's adequacy in capturing the spectral dynamics.

Our findings reveal distinct trends in flux changes across the different regions of the north eastern filament. 
Until 2012, a monotonous increase in flux was observed in both the whole filament (Region 1 and 2) and Region 1. This trend stabilizes between 2012 and 2017, with no significant changes in flux. Conversely, Region 2 exhibited a different pattern, with no discernible increase in flux, highlighting the variability in spectral changes across the filament. 
Intriguingly, a decrease in flux was observed in Region 1 after 2019, suggesting a potential shift in the filament's emission characteristics.

When considering the filament as a whole, the flux increased by more than 50 \% over a span of 12 years, a finding consistent with previous studies (e.g., Patnaude $\&$ Fesen 2009, AJ). 
However, a closer examination reveals that this overall increase is not uniformly distributed across the filament. Specifically, Region 1 experienced an approximate 80 \% increase in flux, 
whereas Region 2 showed no significant change. This difference indicates that the overall flux increase previously reported for the north eastern filament primarily reflects the changes in Region 1, 
with the effects from other regions contributing to an overall variation.

\begin{longtable}{cccccccccc}
  \caption{Best-fit parameters of the spectrum of each region}\label{results}  
\hline\noalign{\vskip3pt} 
    & &  Region 1 and 2 & & & Region 1 & & &Region 2&  \\ 
Year  & Flux$^{*}$ & $\Gamma$ & $\chi^2$/d.o.f.  & Flux$^{\dagger}$ & $\Gamma$ & $\chi^2$/d.o.f. & Flux$^{\dagger}$ & $\Gamma$ & {$\chi^2$/d.o.f.} \\  [2pt] 
\hline
\endfirsthead      
\hline\noalign{\vskip3pt} 
  Name & Value1 & Value2 & Value3 & & & \\  [2pt] 
\hline\noalign{\vskip3pt} 
\endhead
\hline\noalign{\vskip3pt} 
\endfoot
\hline\noalign{\vskip3pt} 
\multicolumn{2}{@{}l@{}}{\hbox to0pt{\parbox{160mm}{\footnotesize
\hangindent6pt\noindent
\hbox to6pt{\footnotemark[$*$]\hss}\unskip%
The flux is given in units of $10^{-12}$ erg $\mathrm{cm^{-2}} \mathrm{s^{-1}}$. 
  {\footnotemark[$\dagger$]\hss}\unskip%
  The flux is given in units of $10^{-13}$ erg $\mathrm{cm^{-2}} \mathrm{s^{-1}}$.
}\hss}} 
\endlastfoot 
2000 & 0.93 $\pm 0.04$  &  3.1 $\pm$ 0.3   & 179.2/116 & 5.12 $\pm 0.30$   & 3.3$\pm$ 0.3 & 151.2/116  & 4.20 $\pm 0.27$   & 2.4 $\pm 0.4$ & 37.3/52\\
2002 & 1.02 $\pm 0.04$ & 2.6 $\pm$ 0.3  & 117.0/116&5.70 $\pm 0.31$ & 2.7$\pm$ 0.3 & 98.0/116&4.33 $\pm 0.27$ &  2.4 $\pm$ 0.4 &41.6/56\\
2004 & 1.08 $\pm 0.01$ & 2.4 $\pm$ 0.1  & 104.7/116&6.50 $\pm 0.07$ & 2.3$\pm$ 0.1 & 102.3/116&4.22 $\pm 0.04$ &  2.5 $\pm$ 0.1 &196.2/116\\
2007 & 1.21 $\pm 0.05$ & 2.2 $\pm$ 0.2  & 112.3/116&7.38 $\pm 0.36$ & 2.2$\pm$ 0.3 & 121.4/116&4.35 $\pm 0.28$ &  2.3 $\pm$ 0.4 &42.6/54\\
2009 & 1.28 $\pm 0.05$ & 2.7 $\pm$ 0.2  & 99.9/116&7.92 $\pm 0.39$ & 2.9$\pm$ 0.3 & 134.1/116&4.29 $\pm 0.29$ &  2.9 $\pm$ 0.4 &41.6/47\\
2010 & 1.31 $\pm 0.05$ & 2.9 $\pm$ 0.2  & 135.3/116&8.27 $\pm 0.38$ & 3.4$\pm$ 0.3 & 140.6/116&4.57 $\pm 0.30$ &  2.3 $\pm$ 0.4 &34.9/52\\
2012 & 1.39 $\pm 0.05$ & 2.3 $\pm$ 0.2  & 126.9/116&8.80 $\pm 0.39$ & 2.7$\pm$ 0.3 & 164.4/116&4.42 $\pm 0.27$ &  2.2 $\pm$ 0.4 &58.1/57\\
2013 & 1.39 $\pm 0.05$ & 2.5 $\pm$ 0.2  & 123.7/116&8.61 $\pm 0.38$ & 3.0$\pm$ 0.3 & 155.1/116&4.71 $\pm 0.29$ &  1.9 $\pm$ 0.4 &63.3/58\\
2014 & 1.38 $\pm 0.05$ & 2.8 $\pm$ 0.2  & 103.7/116&9.00 $\pm 0.39$ & 2.8$\pm$ 0.3 & 127.8/116&4.43 $\pm 0.27$ &  2.8 $\pm$ 0.4 &47.8/57\\
2015 & 1.39 $\pm 0.05$ & 2.6 $\pm$ 0.2  & 144.6/116&9.09 $\pm 0.39$ & 2.5$\pm$ 0.3 & 132.9/116&4.49 $\pm 0.28$ &  3.1 $\pm$ 0.4 &65.4/57\\
2016 & 1.43 $\pm 0.05$ & 2.4 $\pm$ 0.2  & 127.8/116&9.06 $\pm 0.41$ & 2.8$\pm$ 0.3 & 129.0/116&4.89 $\pm 0.31$ &  1.6 $\pm$ 0.4 &58.0/57\\
2017 & 1.38 $\pm 0.05$ & 2.4 $\pm$ 0.2  & 112.7/116&8.95 $\pm 0.39$ & 2.5$\pm$ 0.3 & 126.8/116&4.46 $\pm 0.28$ &  2.4 $\pm$ 0.4 &54.7/57\\
2018 & 1.41 $\pm 0.05$ & 2.7 $\pm$ 0.2  & 86.4/116&9.09 $\pm 0.39$ & 3.1$\pm$ 0.3 & 108.8/116&4.56 $\pm 0.28$ &  2.2 $\pm$ 0.4 &52.9/57\\
2019 & 1.29 $\pm 0.05$ & 3.0 $\pm$ 0.2  & 129.9/116&7.94 $\pm 0.37$ & 3.6$\pm$ 0.3 & 166.7/116&4.31 $\pm 0.28$ &  2.3 $\pm$ 0.4 &61.2/58\\
\end{longtable}

\section{Discussion} \label{sec:dis}

The study of the northeastern filament of Cas~A provides crucial insights into cosmic ray acceleration and magnetic field amplification mechanisms. Building on earlier studies (e.g., Patnaude \& Fesen 2009, AJ), our analysis, which utilizes nearly two decades of \textit{Chandra} observations, identifies an approximately 80 \% increase in X-ray flux on the left side of the filament, while the right side shows minimal change. Despite similar spectral shapes, Region 2 has a slightly lower photon index, consistent with previous findings (\cite{Stage:2006aa}). Morphologically, Region 1 exhibits notable evolution, transitioning from a curved to a straighter configuration, underscoring dynamic changes in its structure.

These results highlight significant spatial variability within the northeastern filament, suggesting that flux increases previously attributed to the filament as a whole are primarily driven by processes in the left region. Furthermore, the flux increase in this region appears to have plateaued over the past decade, offering new constraints on the temporal and spatial dynamics of particle acceleration and magnetic field interactions in SNRs.

A linear approximation of the increase in X-ray intensity in Region 1 from 2000 to 2012 indicates a rate of about 6.0 \%/yr; 
i.e., $(1+ 0.06)^{12}$ is close to 1.8, a ratio of 9 erg s$^{-1}$ cm$^{-2}$ in 2012 to 5 erg s$^{-1}$ cm$^{-2}$ in 2000. 
The acceleration timescale  $\tau_{\rm{acc}}$ (\cite{Sato_2018}) is expressed as: 
\begin{equation}
\tau_{\rm{acc}} \approx \left(25  \text { years}\right) \eta\left(\frac{B}{100~ \mu G}\right)^{-1.5}\left(\frac{\varepsilon}{5~\rm{keV}}\right)^{0.5}\left(\frac{v_{\rm{s}}}{5000~\rm{km/s}}\right)^{-2}, 
\end{equation}
where $\varepsilon$ is a typical synchrotron radiation energy, and $v_{\rm{s}}$ is a shock velocity. 
Given that the Bohm factor, $\eta$, is approximately 1 and $\tau_{\rm{acc}}$ is 20 years,  it yields a magnetic field strength ($B$) about 120~$\mu$G. 
Such an accelerated particles are cooled by radiating synchrotron emission. 
The synchrotron cooling timescale $\tau_{\rm{sync}}$ (\cite{Sato_2018}) is written as follows; 
\begin{equation}
\tau_{\mathrm{sync}} \simeq\left(1.25 \times 10^3 \text { years }\right) E_{\mathrm{TeV}}^{-1} B_{0.1}^{-2},
\end{equation}
where $E_{\mathrm{TeV}}$ is a electron energy, and $B_{0.1}$ is a magnetic field in a unit of 0.1~mG. 
Since the average energy of synchrotron photons is related with $\varepsilon \approx 1.6 B_{0.1} E_{\mathrm{TeV}}^2 \mathrm{keV}$, 
this is interpreted that the electrons within this part of the north eastern filament are being accelerated in the presence of an amplified magnetic field, $\sim$100~$\mu$G. 

Fraschetti et al. (2018) suggest that the time variability of the non-thermal X-ray emission can be attributed to the growth of the magnetic field. This growth is driven by a vortical amplification mechanism occurring when a reflected inward shock collides with inner overdensities.
It is based on an analytical calculation of the vorticity generated downstream using a two-dimensional rippled hydromagnetic shock, neglecting fluid viscosity and resistivity (\cite{PhysRevLett.120.251101, Fraschetti_2013}).
A magnetic field can be saturated up to the order of milli-Gauss, explaining the short-term variability in X-ray observations of SNRs using reasonable parameters for interstellar turbulence. As shown in  Figure \ref{compare}, the morphological changes more apparent in Region 1 may have been caused by collisions with dense clumps, generating vorticity and amplifying the magnetic field in the downstream fluid. This process results in a flux increase that eventually saturates at a certain level. 
In these two regions, no clear distinction has been reported in X-rays; however, the brighter trend in Region 1 has been reported in radio with VLA (\cite{Orlando_2022}) , and the infrared with JWST (\cite{Milisavljevic_2024}). In X-rays, \cite{Sakai_2024} reported that the classification of kinematic properties differs between Region 1 and Region 2.

In these two regions, no significant differences have been reported in X-rays; however, Region 1 appears brighter in radio and infrared (\cite{Milisavljevic_2024}, \cite{Orlando_2022}). 
Additionally,  \cite{Sakai_2024} report that the classification of kinematic properties differs between Region 1 and Region 2. The bright radio emission is likely dependent on synchrotron radiation emitted by electrons moving in a magnetic field, suggesting that Region 1 may have a higher magnetic field. It is also possible that the fluctuations in density and magnetic field are higher in Region 1, and these fluctuations may be reflected in the short-term variability of the X-rays (\cite{Jikei_2024}). To further understand this scenario, deeper observations and analyses, including multi-wavelength data, are needed to further understand this scenario.

In young SNRs such as Cas A, Tycho, Kepler, and SN1006, radial magnetic fields with polarization degrees of a few tens of percent are observed in the radio band. However, the nature of these radial fields has not been thoroughly explored (\cite{10.3389/fspas.2022.882467}).
The global radial magnetic field orientation has been studied through three-dimensional MHD simulations. These studies explain that downstream turbulence, driven by the effects of a rippled shock or Richtmyer-Meshkov instability, is responsible for the observed radial structures of the magnetic field (\cite{Inoue_2013}).
Alternatively, another explanation is that cosmic-ray electrons are selectively accelerated at "locally" quasi-parallel shocks (\cite{West_2017}).
The polarized fraction is a key observable that can help distinguish between different magnetic field scenarios. For an intrinsically radial field, the polarized fraction is very high, but it is significantly reduced when a random turbulent component dominates due to depolarization effects. Recently, X-ray polarization measurements by the X-ray Imaging X-ray Polarimetry Explorer (IXPE) have shown that the measured polarization angle corresponds to a radially oriented magnetic field, similar to what has been inferred from radio observations. However, the X-ray polarization degree is approximately 5 \%, lower than that observed in the radio band (\cite{Vink_2022}).
Since IXPE's angular resolution is $30''$, which is larger than Chandra's $0.5''$, the detailed X-ray polarization map around the north eastern filament remains uncertain and does not appear to be very high. Further investigation is needed to understand the orientation of the magnetic fields in young SNRs, as this is related to particle injection efficiency.

 \begin{figure}[h]
 \includegraphics[width=0.5\textwidth]{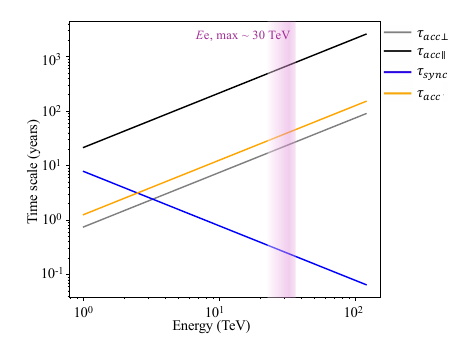}
 \caption{
Acceleration and cooling timescales as a function of electron energy, with the magnetic field set at 120~$\mu$G. The solid gray, black, blue, and orange lines are $\boldsymbol{\tau}_{\mathrm{acc}, \perp}$, $\boldsymbol{\tau}_{\mathrm{acc}, \parallel}$, $\tau_{\mathrm{sync}}$ , and $\tau_{\mathrm{acc}}$, respectively. The purple region indicates the maximum electron energy around 30 TeV, as referenced in Eq.~\ref{Emax}.
{Alt text: Graph with four lines representing timescales and a region showing the maximum electron energy. The x axis shows the energy from 1.0 to 120 tera electron volt. The y axis shows the time scale from 0.03 to 4000 years. }}
 \label{timescale}
 \end{figure}

The acceleration timescale when the magnetic field is parallel to the shock normal direction, $\boldsymbol{\tau}_{\mathrm{acc}, \parallel}$, and when it is perpendicular, $\boldsymbol{\tau}_{\mathrm{acc}, \perp}$, are described by the following equations (\cite{Kamijima_2020}):
\begin{equation}
\boldsymbol{\tau}_{\mathrm{acc}, \parallel} = \frac{r}{r-1}\left[ 1 + r\left(\frac{B_2}{B_1}\right)^{-1} \right]\left( \frac{v_{s}}{\nu} \right)^{-2}\Omega_{\mathrm{g},1}^{-1},
\end{equation}
\begin{equation}
\boldsymbol{\tau}_{\mathrm{acc}, \perp} = \frac{3\pi \eta r}{4\left( r-1 \right)}\left( \frac{v_{\mathrm{s}}}{\nu} \right)^{-1}\Omega_{\mathrm{g},1}^{-1} + \frac{r^2}{r-1}\left( \frac{B_2}{B_1} \right)^{-1}\left( \frac{v_{\mathrm{s}}}{\nu} \right)^{-2}\Omega_{\mathrm{g},1}^{-1},
\end{equation}
where $\Omega_{\mathrm{g}, 1} = q B_1 /(\gamma m c)$ is the gyro angular frequency, and $m$, $q$, $\gamma$,  $B_1$, and c represent the particle mass, charge, Lorentz factor, upstream magnetic field strength, and the speed of light, respectively. $\nu$ is the particle velocity, and $B_2$ is the downstream magnetic field strength.

Figure \ref{timescale} illustrates the acceleration and cooling timescales, assuming a compression ratio $r = 4$, $B_1/B_2 = 1/120$, $v_{\mathrm{s}} = 4.5 \times 10^{3}$ km/s, and a magnetic field of 120~$\mu$G. At lower electron energies, the acceleration timescale is shorter, whereas cooling dominates at higher energies and stronger magnetic fields. These timescales constrain the maximum particle energy, expressed as  :
\begin{equation}
\label{Emax}
E_{\mathrm{e}, \max} \simeq (8.3 \mathrm{TeV}) \frac{4}{\sqrt{3}} \frac{\sqrt{r-1}}{r} k_0^{-1 / 2} B_{0.1}^{-1 / 2} v_{\mathrm{s}, 3},
\end{equation}
where $k_0$ is the ratio $D_0/D_{0\mathrm{Bohm}}$, with $D_0$ and $D_{0\mathrm{Bohm}}$ representing the diffusion coefficient at the electron cutoff energy and at the Bohm limit, respectively. 
$B_{0.1}$ represents the magnetic field in units of $0.1 \, \mathrm{mG}$, and $v_{\mathrm{s},3}$ represents the shock velocity in units of $10^3 \, \mathrm{km/s}$.
This equation is derived from equation (16) of \cite{Parizot2006}, which describes the timescale of particle acceleration in an SNR, and equation (19), which represents the timescale of synchrotron loss in the downstream magnetic field of an SNR.
Substituting $r = 4$, $k_0 = 1$, $B = 120\,\mu\mathrm{G}$, and $v_{\mathrm{s}} = 4.5 \times 10^3$ km/s into equation (5) gives $E_{e, \max} \sim 34$ TeV. This suggests that, for typical values of the shock velocity and magnetic field, the maximum energy is approximately 30 TeV. Therefore, we have indicated this region in Fig.~4 as a reference. However, the estimated maximum electron energy depends on the assumed parameter values and the equation used. 
The perpendicular shock's acceleration timescale is about 20 years, comparable to the flux changes observed in Region~1. Conversely, the parallel shock exhibits a longer acceleration timescale. Differences in the timescales between Region  1 and Region 2 may stem from variations in magnetic field orientation.


According to \cite{Caprioli_2014}, in quasi-parallel configurations -- where the background magnetic field is nearly aligned with the shock normal -- DSA operates efficiently, leading to significant particle acceleration, particularly at large Mach numbers. In contrast, the acceleration efficiency decreases in nearly perpendicular shocks. Moreover, the generation of magnetic turbulence, which is closely associated with efficient ion acceleration, is suppressed in quasi-perpendicular configurations. If the angle between the background magnetic field and the shock surface differs between Region 1 and Region 2, reflecting the turbulent structure downstream of the shock, this could appear as variations in X-ray polarization, which may be detectable in future X-ray polarization observations.

Beyond analytical methods, stochastic shock drift acceleration (SSDA), shock surfing acceleration (SSA), and magnetic reconnection should be considered as electron acceleration mechanisms (\cite{2022RvMPP...6...29A}). Advanced observations, such as those enabled by XRISM (\cite{10.1117/12.3019325}) and future instruments with higher spatial and polarimetric resolutions (\cite{Kamijima_2020, 10.1117/12.3019325}), are crucial for understanding these processes. Multidimensional fully kinetic Particle-in-Cell simulations (\cite{PhysRevLett.119.105101}) reveal that electrons are initially energized by SSA at the shock's leading edge before undergoing further acceleration by SSDA within the deeper shock transition layer. Such theoretical predictions must be validated through observations, accounting for environmental effects in regions like Cas~A.

Recent results from NuSTAR, Fermi-LAT, VERITAS, and radio observations indicate that the classical DSA model alone cannot explain the over-a-decade-long spectral energy distribution and its dimming trend. A hybrid model, proposing two cosmic-ray electron populations -- an older, softer component and a recently accelerated, harder component -- has been suggested (\cite{Woo:2024sxw}). However, direct verification remains challenging due to the spatial resolution of gamma-ray observations being insufficient for \textit{Chandra} . In SNRs with complex structures, detecting localized, short-term variations with high spatial resolution will be essential for advancing our understanding of these processes.

Regarding the origin of the NE filament, as higher Mach numbers generally favor particle acceleration at forward shocks, the most efficient acceleration regions propagate outward as shown in the simulation of SN 1006 (\cite{Huan_2015}). 
However, the NE filament in Cas~A, despite being close to the Bohm limit, is not located at the outermost edge. This may relate to recent studies indicating that magnetic field generation and amplification occur within a Mach number range where only electrons are magnetized (\cite{Jikei_2024}). Observations in radio, optical, and near-infrared suggest that Region 1 exhibits brightening, implying higher density and magnetic field fluctuations, which could lead to differences in the Alfv\'{e}n Mach number and explain the observed short-term X-ray variations. Further investigation is required to understand the interplay between the pre-explosion environment and explosion asymmetry.

\section{Summary} \label{sec:sum}

This study investigates the time evolution of the northeastern filament of Cas~A, focusing on its localized morphological and spectral changes. Over two decades of \textit{Chandra} observations, we identified significant morphological changes and an 80\% increase in non-thermal emission flux in one section of the filament, which eventually stabilized. In contrast, the other section showed minimal flux and morphological changes. These results suggest that particle acceleration and magnetic field amplification are highly localized processes. The findings provide critical insights into the dynamics of supernova remnants and their role in cosmic ray origins. Long-term observations by \textit{Chandra}, complemented by \textit{NuSTAR}, IXPE, and XRISM, will be essential in mapping these processes. Future detailed observations with high spatial resolution and multi-wavelength studies will be a key in understanding the asymmetric nature of ejecta and the complex interactions of magnetic fields within Cas~A.


\begin{ack}
This work was supported by JSPS KAKENHI Grant Numbers 18H03722, 19K14749, 20H01941, 20K20527,  22H01272, and 24K00672. 
\end{ack}

\appendix 
\section*{Appendix1. Details on the spectral comparison} \label{sec:spec}

\begin{figure*}[h]
 \includegraphics[width=0.90\textwidth]{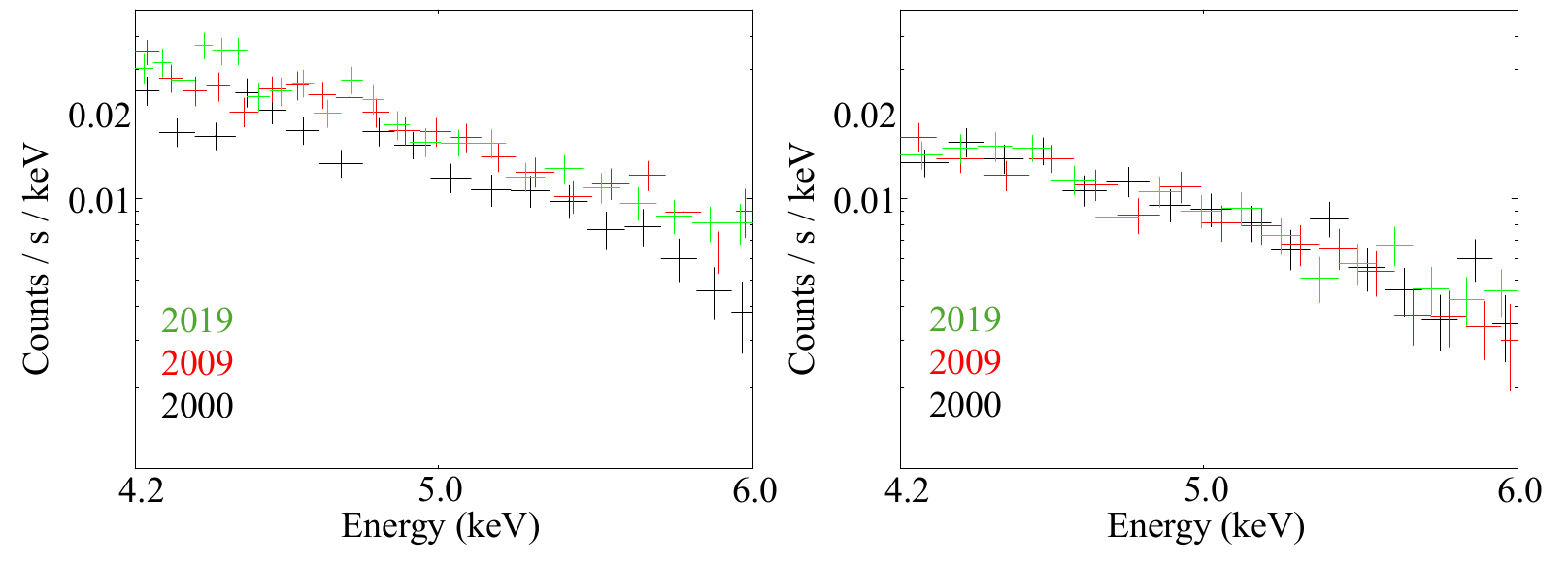}
 \caption{(Left) The count rate spectra of Cas~A of Chandra, obtained from OBSID 114 (2000), OBSID 12020 (2009), and OBSID 19606 (2019) for Region 1. (Right) Corresponding spectra for Region 2 across the same observational IDs.
 {Alt text: The two panels have scatter plots of fluxes for 2000, 2009, and 2019, respectively. In both graphs, the x axis shows energys from 4.2 to 6.0 kilo electron volt  and the y axis shows the number of counts per second per kilo electron volt from 0.002 to 0.03. } } 
 \label{compspec}
 \end{figure*}

For each observation, the source and background regions are selected as illustrated in Figure~\ref{CasAfil}. The spectra derived from these regions, specifically Region 1 and Region 2, are shown in Figure~\ref{compspec}. To facilitate straightforward visual comparison, we have rebinned the spectral data tightly, ensuring that the flux levels of each spectrum are discernible by eye.

The data shown are collected in the years 2000, 2009, and 2019. Analysis of the spectral evolution in Region 1 reveals a noticeable trend of gradual brightening starting in 2000, with the rate of increase appearing to plateau between 2009 and 2019. In contrast, the spectra from Region 2 show little change in flux levels throughout the two-decade span of our study. 
The raw spectra also indicates that Region 1 has a slightly steeper spectral slope, 
which is consistent with the fitting results shown in table \ref{results}. 

\section*{Appendix2. Systematic uncertainties in the brightening of the non-thermal component} \label{Apendix2}
We provide supplementary information on the two methods introduced in Section 3.2 and the corresponding results.
The first method is considered suitable for capturing the overall trend of the flux increase. We defined a source region that adequately covers the northeastern filament, and selected upstream and downstream regions relative to the explosion center -- where thermal emission is expected to dominate -- as background regions. By subtracting the spectra of these thermal-dominated regions from the source spectrum, we obtained a difference spectrum. 
As a reference, Figure 6 shows an example of the source spectrum from the year 2000, the background (thermal-dominated) spectrum, and the resulting difference spectrum. 
Since the spatial variation of the thermal component is generally small, the difference spectrum can be reasonably approximated by a power-law model. 
We fitted the difference spectra for all epochs using a power-law model, and the temporal evolution of the derived fluxes is shown in Figure 7. 
Although the absolute flux values differ by approximately 20~\% from those presented in Figure 3 due to systematic uncertainties, the trend of flux increase by a factor of ~2 remains significant.

In the second method, we defined a background region far from the source to avoid contamination from astrophysical signals, and performed broadband spectral fitting of the northeastern filament. 
For this, we adopted the model tbabs(vpshock + power) in XSPEC notation and applied it to all epochs. 
The fluxes of the power-law component derived from this method are also shown in Figure 7. 
As with the first method, the absolute fluxes differ from those in Figure 3 by approximately 20~\%, which we attribute to systematic uncertainties. Nonetheless, the observed brightening by a factor of ~2 is considered significant.
 \begin{figure*}[h]
 \includegraphics[width=\textwidth]{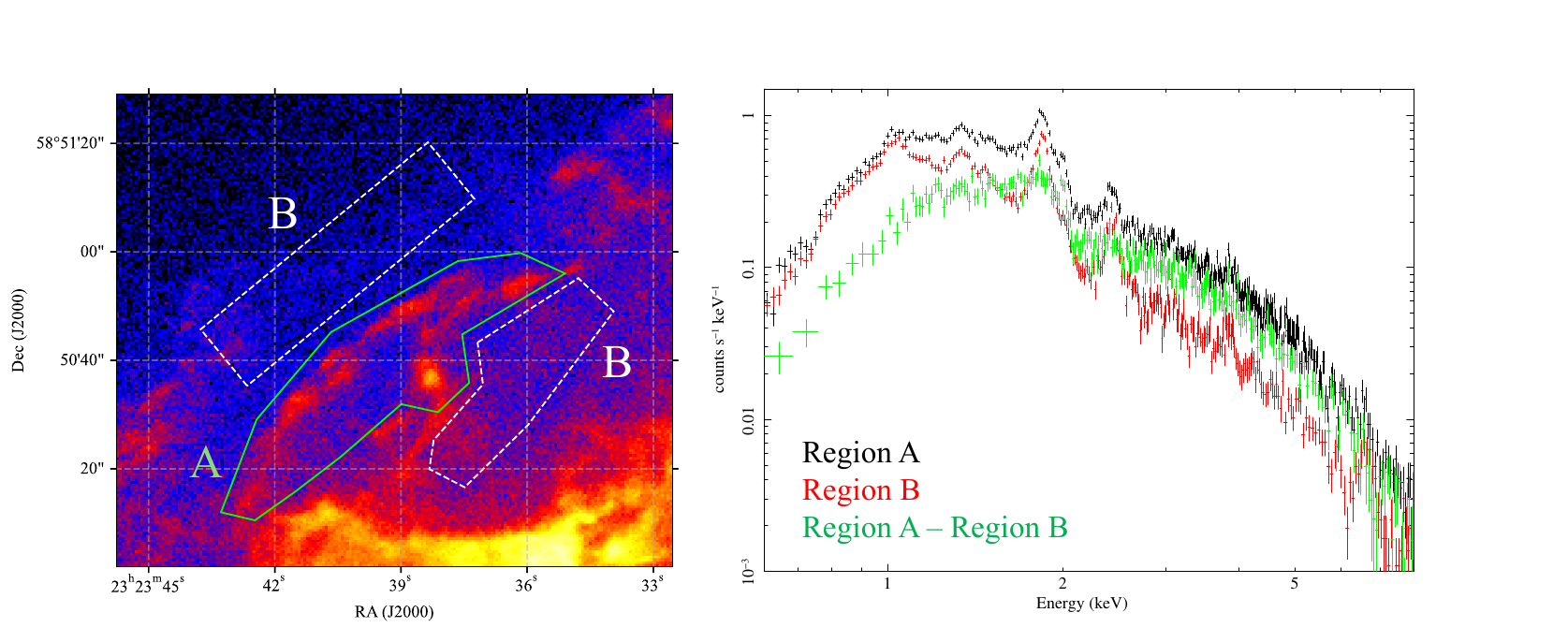}
 \caption{(Left) Analysis regions of method A. The central region A is the filament, and regions B on both sides are backgrounds. (Right) Spectral analysis results for method A. The spectra of region A, regions B, and their difference are shown in black, red, and green, respectively. {Alt text: The left panel displays the analysis regions. The filament is enclosed within a polygon, defined as region A. Polygonal background regions B are located just inside and outside the filament. The right panel presents the spectra of each region, along with their difference. The spectrum of region A includes both thermal and non-thermal components, whereas the spectrum of regions B is primarily thermal. The difference spectrum highlights the non-thermal component by subtracting the thermal contribution.} } 
 \label{methodAB}
 \end{figure*}
 
  \begin{figure}[h]
 \includegraphics[width=0.45\textwidth]{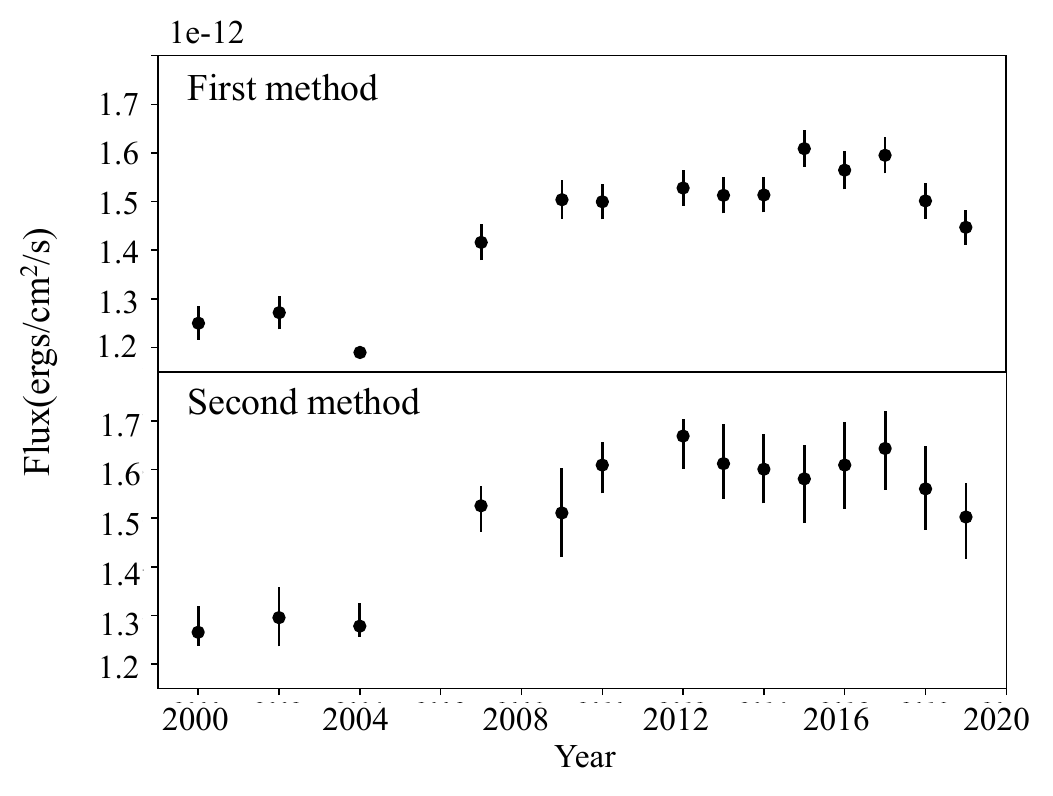}
 \caption{Flux variations obtained by the first method and the second method. The upper panel shows results from the first method, and the lower panel shows results from the second method. {Alt text: Two scatter plot graphs are shown. In both graphs, the x-axis represents the observation year, ranging from 2000 to 2020. In the upper panel, the y-axis indicates the flux derived using the first method. In the lower panel, the y-axis shows the flux obtained using the second method. For both methods, the flux increased significantly until around 2012, similar to the trend observed with the third method described in Section 3.2. } } 
 \label{methodAB}
 \end{figure}

\bibliography{CasA_ronbun_kominato_20250418}

\bibliographystyle{aasjournal}

\end{document}